\documentclass[a4paper,11pt]{article}

\usepackage{jinstpub} 
\usepackage{lineno}
\usepackage{placeins}
\usepackage{soul}

\usepackage{titlesec}
\usepackage{subfigure}
\usepackage[normalem]{ulem}
\usepackage{graphicx}
\usepackage{subcaption}
\usepackage{cleveref}

\title{\boldmath Beam Test of a SiPM-on-Tile ZDC Prototype with 5.3 GeV Positrons at Jefferson Laboratory }

\author[a]{Sean Preins,}
\author[a]{Weibin Zhang,}
\author[a]{Ryan Tsiao,}
\author[a]{Mia Macias,}
\author[a]{Brice Saunders,}
\author[c]{Love Preet,}
\author[a, b]{Miguel Arratia}

\affiliation[a]{Department of Physics and Astronomy, University of California, Riverside, CA 92521, USA}
\affiliation[b]{Thomas Jefferson National Accelerator Facility, Newport News, Virginia 23606, USA}
\affiliation[c]{Department of Physics, University of Regina, Saskatchewan S4S 0A2, Canada}
\date{}

\emailAdd{miguel.arratia@ucr.edu}

\abstract{We report on a beam test of a Silicon Photo-Multiplier (SiPM)-on-tile Zero Degree Calorimeter (ZDC) prototype developed for the future Electron--Ion Collider (EIC). The detector implements the staggered scintillator-tile geometry envisioned for the final detector and includes 370 instrumented channels, corresponding to $\mathcal{O}(10\%)$ of the full ZDC. The detector was tested using a 5.3~GeV positron beam at Jefferson Laboratory. We measure the energy response, shower shape, and spatial reconstruction performance of the detector and compare these results with simulation. These studies provide key input for the optimization of the final ePIC ZDC design.}


\begin{document}

\maketitle
\flushbottom

\section{Introduction}
A key requirement for the physics program of the future Electron--Ion Collider (EIC) \cite{Accardi2016, ABDULKHALEK2022122447} is the precise detection of neutral particles emitted at very small angles with respect to the hadron beam. Within the ePIC detector \cite{ePIC}, the Zero Degree Calorimeter (ZDC) ~\cite{MILTON2025170613} plays a central role in this capability, enabling measurements of forward neutrons and photons. To meet these requirements, the ePIC ZDC is designed as a compact, high-granularity sampling calorimeter based on Silicon Photo-Multiplier (SiPM) on-tile technology. Its modular architecture allows access to sampling layers for periodic SiPM annealing, mitigating radiation-induced degradation \cite{UCDavisTest}.

SiPM-on-tile calorimetry employs plastic scintillator tiles coupled to individually instrumented SiPMs, enabling high transverse granularity and adaptable mechanical integration. These characteristics make the technology well suited for highly granular calorimeters and have motivated its adoption in several modern detector systems. Implementations have been studied extensively by the CALICE collaboration~\cite{RevModPhys.88.015003} and incorporated into the CMS High-Luminosity LHC calorimeter upgrade~\cite{CERN:HL-LHC, Contardo:2015_phase2, CMS:2017_HGCAL_TDR}. Within ePIC, related SiPM-on-tile designs are planned for multiple subsystems, including the high-granularity calorimeter insert~\cite{ARRATIA2023167866} and the forward and backward hadronic calorimeters~\cite{BOCK2023168464}, in addition to the ZDC. These design features were originally motivated by the requirements of particle-flow calorimetry~\cite{THOMSON200925} for proposed lepton collider detectors~\cite{theildconceptgroup2010internationallargedetectorletter, theildcollaboration2020internationallargedetectorinterim, linssen2012physicsdetectorsclicclic, FCC_ee_CDR_Vol2, thecepcstudygroup2018cepcconceptualdesignreport}. Validating this technology across the wide range of operating conditions expected at the EIC is a high priority for the experiment.

Operating at pseudorapidities $\eta > 6$, the ePIC ZDC must achieve high energy and position resolution for electromagnetic and hadronic showers while remaining compact and accessible for servicing. The ZDC design employs an iron--scintillator sampling structure with small scintillator tiles instrumented with SiPMs, arranged in a staggered geometry that enhances transverse position resolution~\cite{MILTON2025170613,Paul_2024}. Demonstrating that this configuration meets performance expectations in beam conditions is a critical step toward final detector construction.

This work builds on a sequence of previous SiPM-on-tile research and development efforts, including bench-top characterization of scintillator tiles and SiPMs~\cite{Arratia_2023}, a 40-channel beam test conducted at Jefferson Laboratory~\cite{instruments7040043}, and a 192-channel prototype deployed in the STAR experimental hall at RHIC~\cite{Zhang_2025}. Together, these studies demonstrated the feasibility and scalability of the SiPM-on-tile approach in both controlled test-beam environments and collider-like conditions. After the measurements presented in this paper were completed, the same detector was exposed to radiation equivalent to one year of operation at the EIC, and it was shown that the detector could still be calibrated on a channel-by-channel basis under realistic radiation damage conditions~\cite{Zhang:2025cmc}.

We report a beam test of a 370-channel SiPM-on-tile ZDC prototype conducted at Jefferson Lab’s Hall D Pair Spectrometer in April 2025. The detector represents a significant scale-up relative to earlier prototypes, comprising 370 readout channels and implementing the staggered scintillator layout foreseen for the final detector. The test was conducted in the Hall~D Pair Spectrometer (PS) using positrons with energies of approximately 5.3~GeV and accumulated 6.58 million events over the course of the run, with 98.7\% of channels fully operational. The objectives of this test were to validate the ZDC detector concept and geometry, benchmark the detector response against simulations, and assess key performance characteristics, including energy resolution, spatial resolution, and shower shape. This study also provided operational experience with a ZDC-specific implementation of SiPM-on-tile technology and informs the ongoing optimization of the final ePIC ZDC design.

\FloatBarrier
\section{ZDC Prototype}
\label{sec:testmodule}

\subsection{Overview of Structure}
\begin{figure}[h!]
    \centering
        \includegraphics[width=0.8\linewidth]{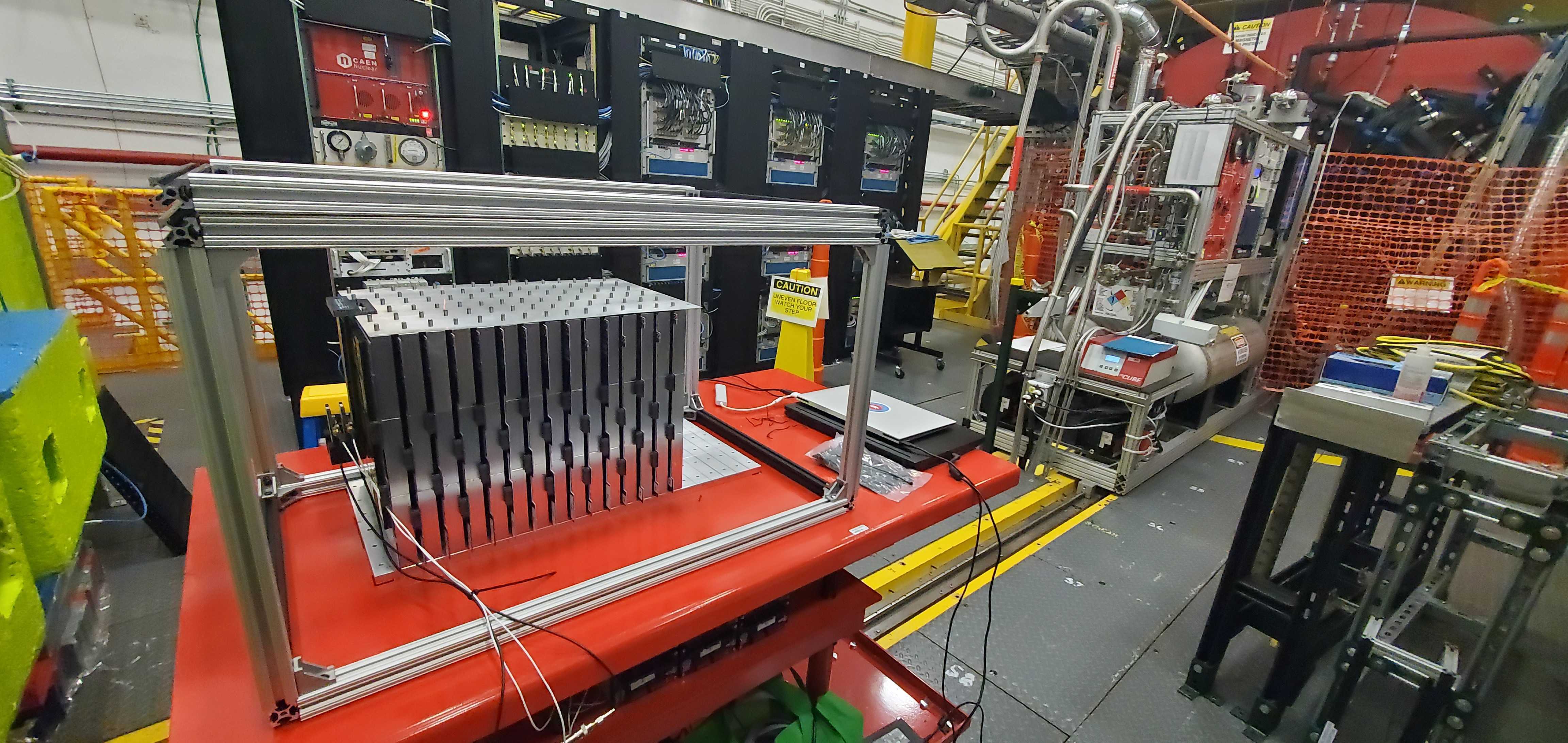}
        \caption{The ZDC prototype installed at Jefferson Lab in front of the Hall-D Pair Spectrometer}
        \label{fig:install}
\end{figure}

The ZDC prototype (Fig.~\ref{fig:install}), installed at Jefferson Lab in front of the Hall-D Pair Spectrometer, is a 15-layer SiPM-on-Tile sampling hadronic calorimeter. Each layer consists of a $5 \times 5$ array of square \( 48.8\,\text{mm} \times 48.8\,\text{mm} \times 4\,\text{mm} \) EJ-212 scintillating tiles, individually wrapped in reflective ESR foil and mounted on a custom PCB. Hamamatsu s14160-1315PS 1.3-mm SiPMs are soldered to the PCB, and the scintillating tiles are air-coupled to the SiPMs via a central dimple. The PCB routes the signals to adapter connectors that attach to the DAQ system via 28 AWG ribbon cables.

As individual layers were being constructed, a series of quality assurance (QA) checks were performed to verify proper SiPM operation and signal integrity before integration into the full detector stack.

The layers are interleaved with 20 mm thick steel plates (1.1 $X_0$), with alternating layers shifted diagonally by half a cell width in each direction. In total, the detector was approximately 17 $X_0$ long, and had the dimensions \( 28.7\,\text{cm} \times 29.5\,\text{cm} \times 61.0\,\text{cm} \).
The entire detector is housed within an 8020-bar frame wrapped with black-out fabric to act as a dark box. The DAQ system is set up on a platform beneath the detector.

\subsection{DAQ and Trigger System}

The ribbon cables from the PCB layers are connected to six CAEN DT5202 units, linked via TDLink fiber optic cables to a CAEN DT5215 Concentrator Board, which connects to a DAQ laptop via USB, controlled remotely through RealVNC.

\begin{figure}[h!]
    \centering
        \includegraphics[width=0.9\linewidth]{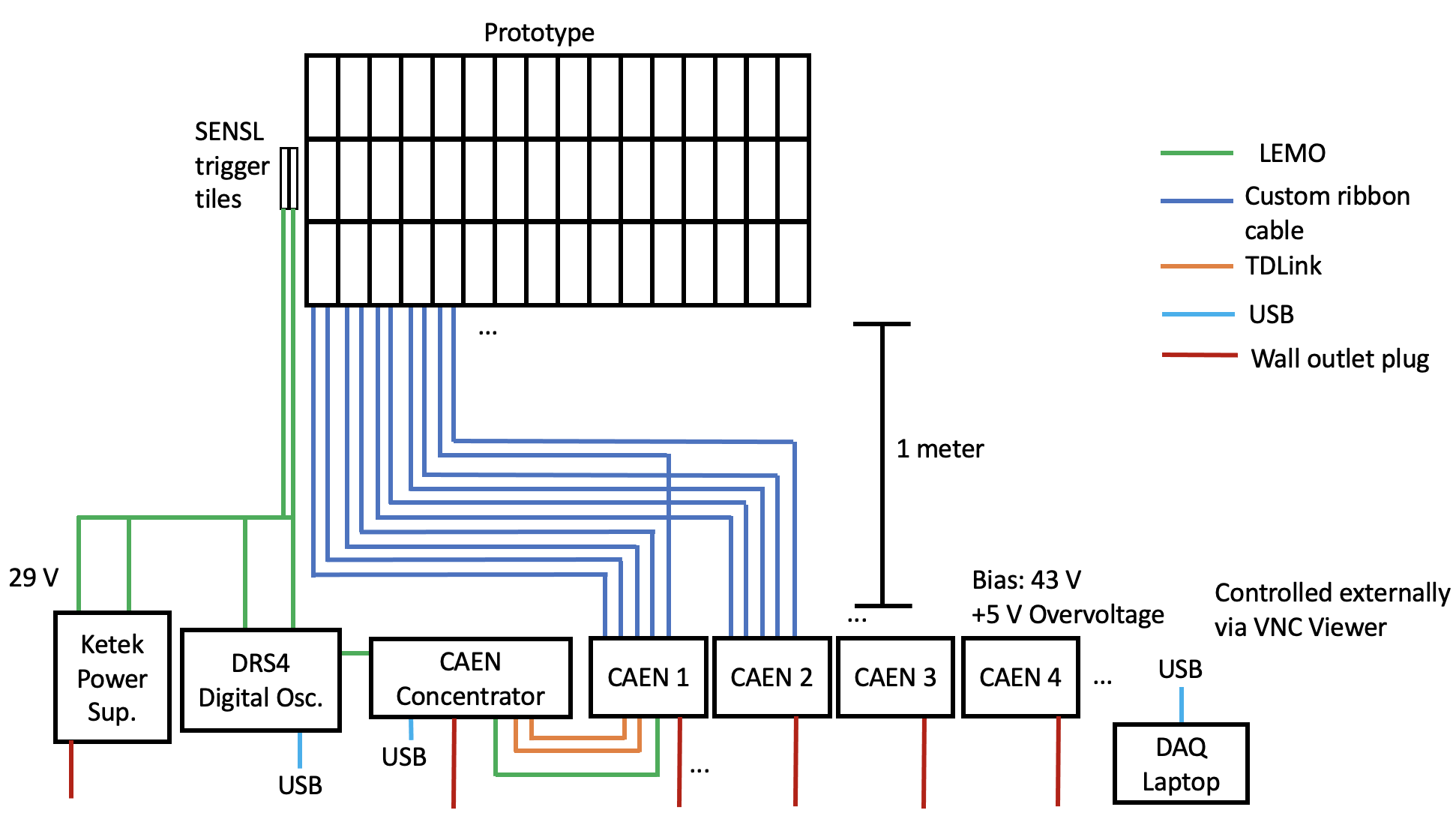}
        \caption{A diagram of the DAQ system for the ZDC prototype.}
        \label{fig:daq}
\end{figure}

The DRS4 trigger output is fed into an F\_IN port on the concentrator, duplicated across six F\_OUT ports. These trigger signals are distributed to the T0-IN ports of the six CAEN DT5202 units via LEMO cables. A diagram of the entire DAQ system is shown in Fig.~\ref{fig:daq}.

\begin{figure}[h!]
    \centering
        \includegraphics[width=0.3\linewidth]{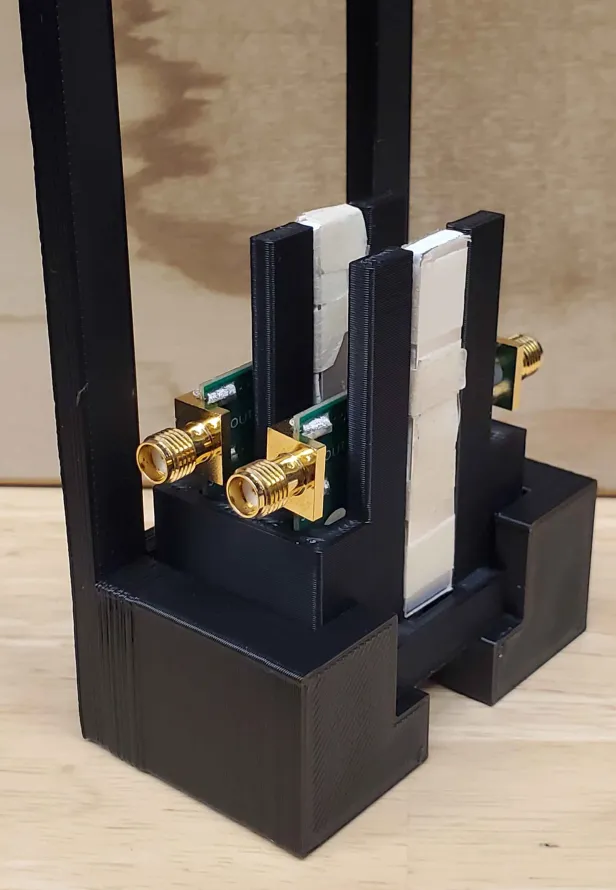}
        \caption{The trigger system for the ZDC prototype.}
        \label{fig:trigger}
\end{figure}

The trigger system consists of two SENSL boards with 3-mm SiPMs, air-coupled to thin scintillating tiles aligned in-line. These are mounted in a 3D-printed holder shown in Fig.~\ref{fig:trigger}, suspended from the iron structure's top pins. The SENSL boards are biased at 29 V using a KETEK SiPM bias source in parallel. The SOUT signal is transmitted via SMA cables to a DRS4 digital multimeter, connected to a laptop via USB and controlled remotely through RealVNC.

The DRS4 software is configured for AND logic between the two channels with a threshold of 5 mV (0.17 MIP).

\FloatBarrier
\section{Commissioning and Data Acquisition}
\label{sec:setup}
\subsection{Installation}

The ZDC prototype was installed at the Hall-D PS at JLab on April 23, 2025. After a series of QA checks, 370 of the total 375 channels were determined to be functional. All non-functional channels were located outside of the central shower region of the detector. The detector was installed on the positron side of the Hall-D PS, $0.09$ radians away from the beamline, oriented radially from the diverging point within the Hall-D dipole magnet. The center of the detector's front layer was positioned 438.8 cm downstream, 39.4 cm right of the beamline, and level with it. The position and orientation were determined with the aid of a laser level device. The detector operated here from April 23 - 30, 2025, receiving 5.3 GeV positrons from the Hall-D PS ~\cite{somov2025private} intermittently at variable intensity.  

Beam-on data were collected from the detector using the JANUS DAQ software. For the results presented in this analysis, data were taken using a single operating configuration selected to optimize the usable dynamic range of both the high- and low-gain readout channels. JANUS was set to recorded in SPECTROSCOPY mode, with the pedestal position set to 100 ADC. The SiPMs were biased at 43~V, and the front-end electronics were configured with a high-gain setting of 63 and a low-gain setting of 58. A hold delay of 100~ns and a low-gain shaping time of 25~ns were used, and triggers were provided via the external T0-IN input.

Prior to beam-on data taking, short PTRG (periodic trigger) runs were performed to characterize the pedestal response of each channel under the same operating conditions. The mean pedestal values determined from these runs were subtracted from each corresponding beam-on hit. To suppress electronic noise, channel entries with amplitudes below three times the pedestal width (\(3\sigma\)) were set to zero. This configuration yielded approximately \(6.58\times10^{6}\) beam-on events and was used for all results presented in this work.

\subsection{Channel Equalization}

Before beam delivery, the detector recorded approximately one day of cosmic-ray data to perform a channel equalization using minimum ionizing particles (MIPs). Events were collected with the same electronics configuration as during beam operation, and required a coincidence of two channels above threshold. The MIP signals from cosmic rays were used to determine relative gain corrections between channels.

For each channel, the high-gain (HG) energy spectrum was fit with a cubic spline, and the most probable value above the pedestal region was extracted as the reference response. Beam-on data taken with identical settings were then used to determine the low-gain to high-gain (LG/HG) ratio for each channel, allowing LG signals to be placed on a consistent relative scale across channels.

Because trigger threshold effects were found to distort the observed MIP spectrum, this procedure was used only to equalize the relative channel gains. A separate global calibration factor was subsequently applied to align the mean event energy with the nominal beam energy.

\subsection{Event Building}

\begin{figure}[h!]
    \centering
    \includegraphics[width=0.49\linewidth]{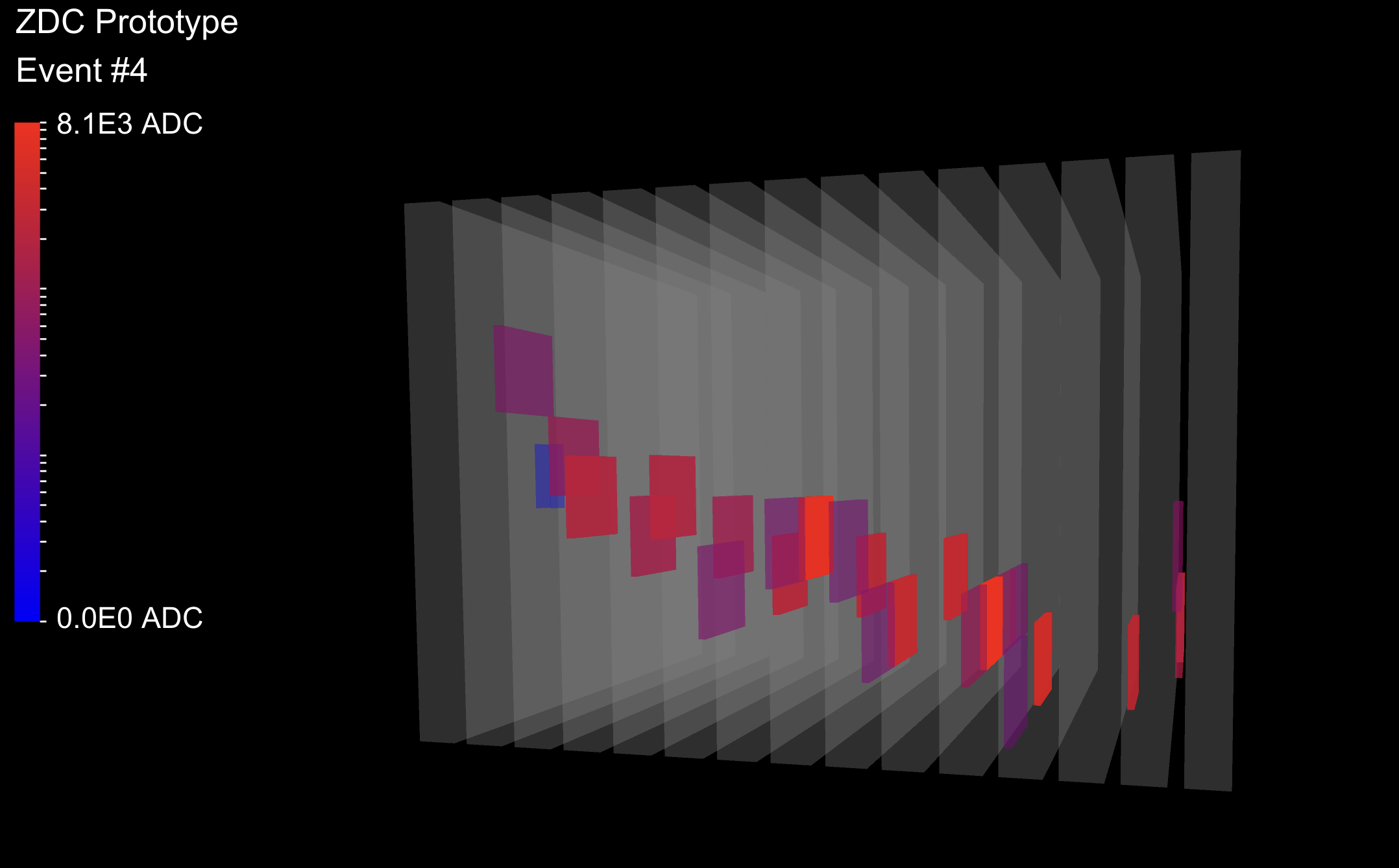}
    \includegraphics[width=0.49\linewidth]{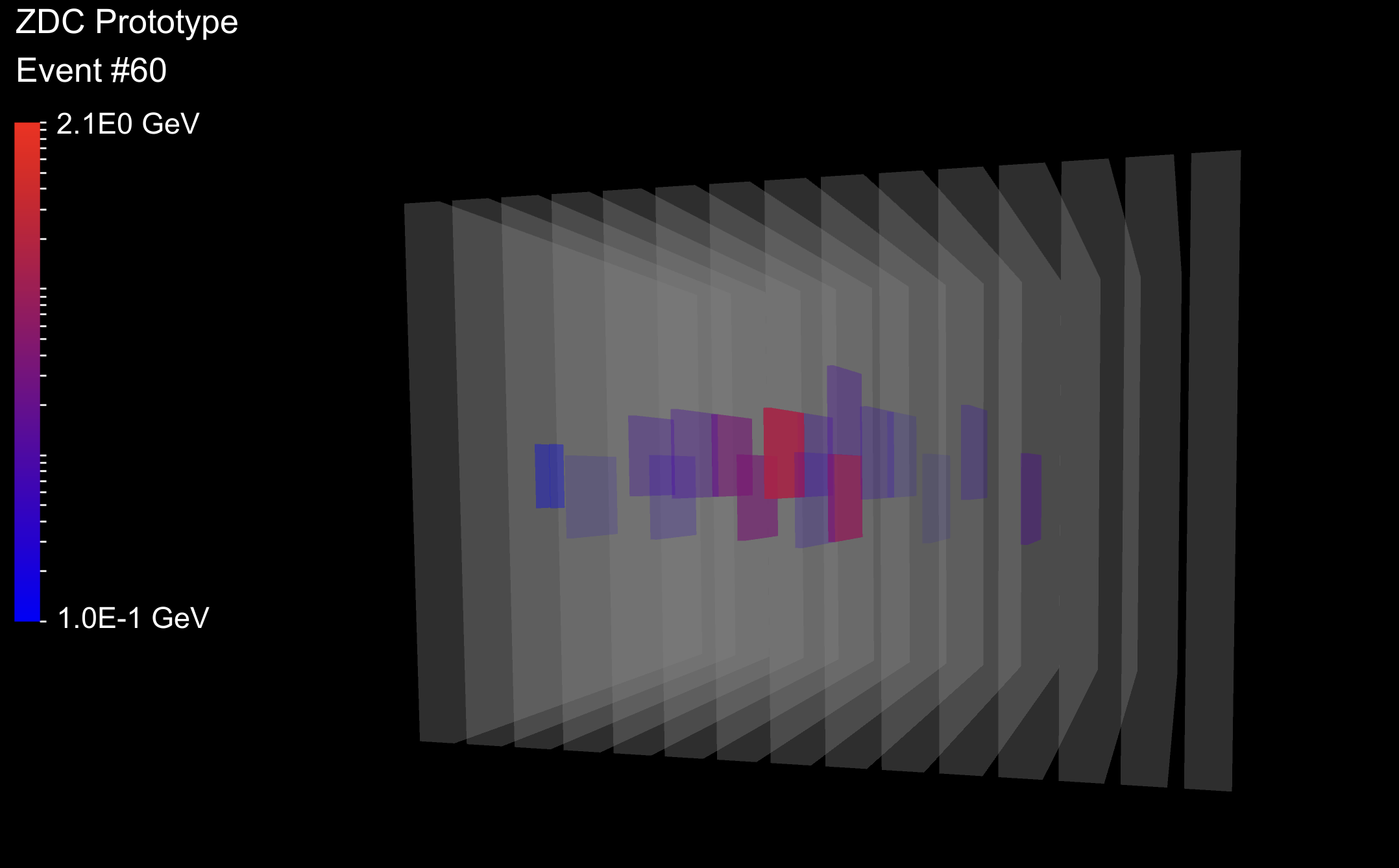}\\
    \caption{A typical cosmic ray event (left) and beam-on event (right).}
    \label{fig:event_display}
\end{figure}

A custom script was used to synchronize the events recorded by each individual DT5202 CAEN unit using the recorded time stamp for each event. To confirm event building was functioning as expected, the ZDC prototype was set to record cosmic ray events using the external triggers while the beam was off. A visualization of a typical cosmic ray event is shown in Fig. \ref{fig:event_display}. As expected, the reconstructed cosmic ray track intersects with the middle front of the detector, where the trigger tiles are located. This also serves as a check to ensure that the channel mapping and geometry have been encoded properly. A typical reconstructed beam-on event is also shown in Fig. \ref{fig:event_display}.

\FloatBarrier
\section{Results}
\label{sec:results}

\subsection{Simulation and Event Selection}
After the events from a chosen JANUS setting configuration were event built and the equalization procedure was performed, the results were compared with a simulation of the ZDC prototype using the GEANT4 framework, with $5.3$~GeV positrons incident near the front center of the detector. The simulated beam was angled at $\theta = 0.07$~rad and $\phi = -0.52$~rad and positioned at $x = -1.2$~mm and $y = -10$~mm to match the measured beam conditions (Section~\ref{sec:position}). In total, $8.9 \times 10^{5}$ events were simulated.

For both data and simulation, a hit energy cut of $E > 0.24$~MeV was selected, and then all hit energies were scaled using a sampling fraction determined from simulation. A cut on events with a total energy less than $1$~GeV and a hit multiplicity less than $10$ was applied to both the simulation and data, which removed $9.6\%$ of the events from the data, and no events from the simulation.

\subsection{Systematic Errors}

Systematic uncertainties were estimated by generating alternative hit energy datasets for each source of uncertainty and computing event-level observables independently. For each source, variations in histogram bins were combined in quadrature to obtain the total bin-by-bin systematic uncertainty.

One source of systematic uncertainty is the detector position, known to approximately $\pm 1.3$~cm relative to the Hall~D beamline. Given the Hall~D PS energy spread of 36.7~GeV/rad \cite{BARBOSA2015376}, this translates to an uncertainty of $\pm 0.1$~GeV at 5.3~GeV, or a relative uncertainty of $\pm 2.0\%$. To evaluate the impact of this uncertainty, two alternative hit energy datasets were constructed, one where all hit energies were scaled by a factor of $0.98$ and one scaled by a factor of $1.02$. These variations resulted in no significant impact on the energy or position resolutions.

An additional systematic uncertainty arises from the channel equalization procedure. In the nominal reconstruction, each channel is equalized using its measured MIP scale in the high-gain readout, together with its individual low-gain to high-gain (LG/HG) ratio determined from beam-on data. To quantify the systematic uncertainty this procedure introduces, an alternative reconstruction was produced in which all channels were calibrated uniformly. This variation produced a $-0.9\%$ impact on the energy resolution, a $-2.1\%$ impact on the $y$ position resolution, and a $+8.1\%$ impact on the $x$ position resolution.

\subsection{Detector Performance}

\subsubsection{Position and Angle Distributions}
\label{sec:position}
For each shower, the moment matrix is defined by
\begin{equation}
M = \frac{1}{\sum_i w_i} \sum_i w_i (\vec{r}_i - \vec{r}_{\text{CoG}})(\vec{r}_i - \vec{r}_{\text{CoG}})^T
\end{equation}
Where $w_i$ is the normalized energy weight of hit $i$, $\vec{r}_i$ is the location of hit $i$, and $\vec{r}_{\text{CoG}}$ is the location of the center of gravity of the particle shower.

\begin{figure}[h!]
    \centering
        \includegraphics[width=0.99\linewidth]{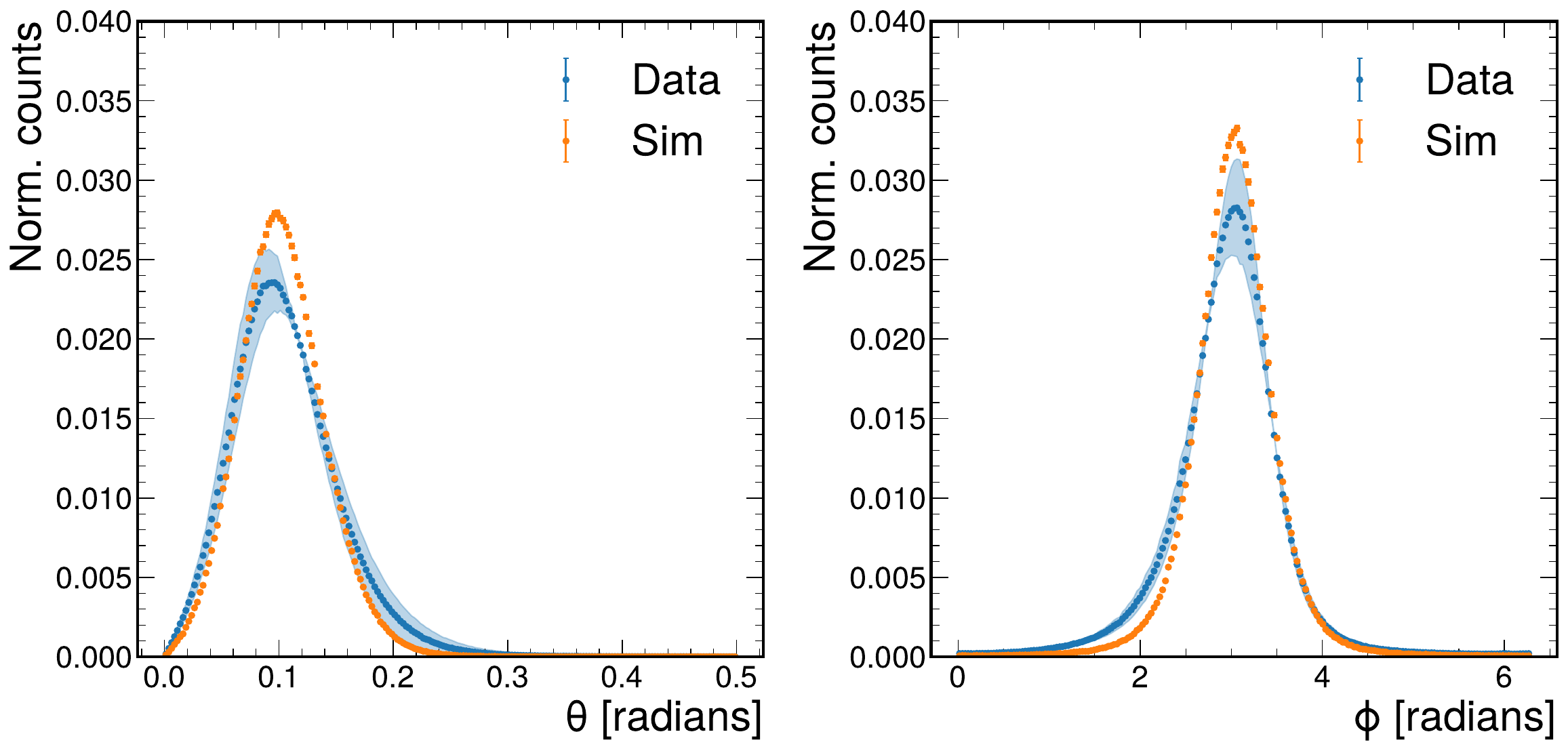}
        \caption{Angular distributions of particle showers compared to a simulation of 5.3 GeV positrons. The combined systematic uncertainty is represented as an error band, and statistical uncertainty is represented with error bars.}
        \label{fig:angular_dist}
\end{figure}

The eigenvectors of $M$ define the principal axes of the shower, which provide sensitivity to the incident angle of the incoming particle. The angular distribution of the principal axes was used to determine the simulated particle gun angle relative to the detector, yielding $\theta = 0.07$~rad and $\phi = -0.52$~rad (Fig.~\ref{fig:angular_dist}).

\begin{figure}[h!]
    \centering
        \includegraphics[width=0.99\linewidth]{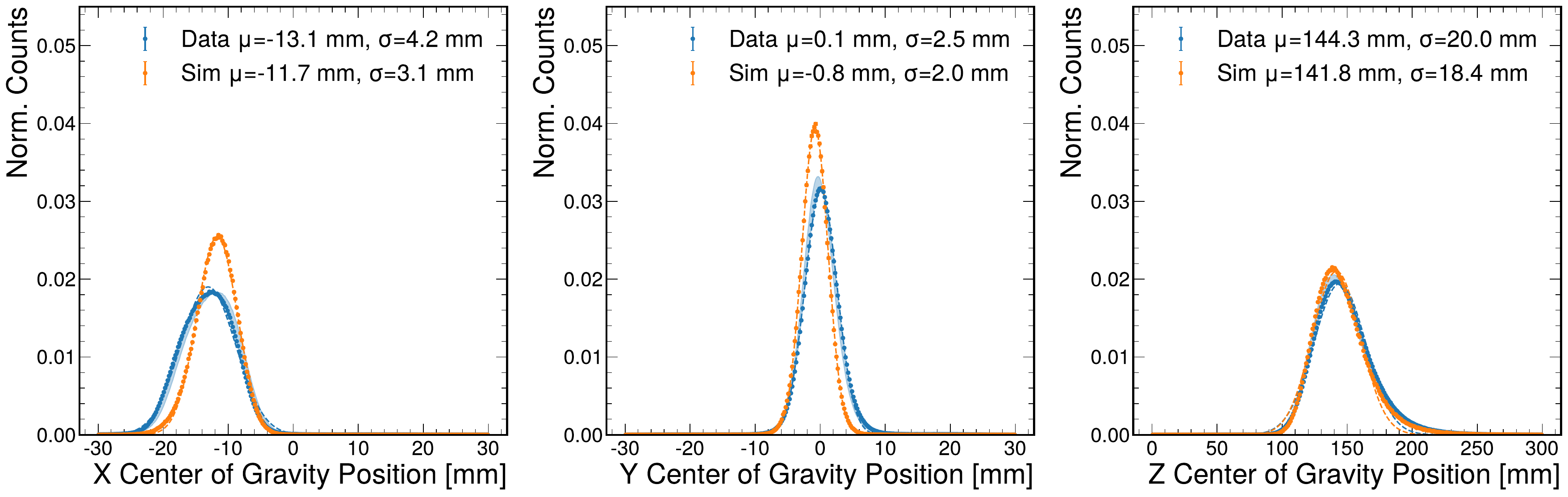}
        \caption{Shower center of gravity distributions compared to a simulation of 5.3 GeV positrons. The combined systematic uncertainty is represented as an error band, and statistical uncertainty is represented with error bars.}
        \label{fig:cog_distributions}
\end{figure}

The shower center of gravity distributions were used to determine the simulated particle gun position at the front end of the detector, which was set at $x = -1.2$~mm and $y = -10$~mm (Fig.~\ref{fig:cog_distributions}). The beam impact position for each shower was then calculated by projecting the principal-axis vectors from the center of gravity to the front surface of the detector. The reconstruction algorithm exhibits systematic offsets between reconstructed and true angles and positions due to tile granularity and detector element offsets.

In the horizontal ($x$) direction, the beam width is constrained by the trigger tile of width 10.6~mm, corresponding to an RMS of $10.6 / \sqrt{12} = 3.1$~mm, assuming a uniform distribution. In the vertical ($y$) direction, the beam is defined by a 5~mm collimator, producing a uniform distribution at the detector corresponding to an RMS of 1.44~mm. These contributions were subtracted in quadrature from the measured RMS of the position distribution to obtain the intrinsic detector resolution in data. As no beam spread effects were included in the simulation, no subtraction in quadrature was applied to the simulation results. 

\begin{figure}[h!]
    \centering
        \includegraphics[width=0.99\linewidth]{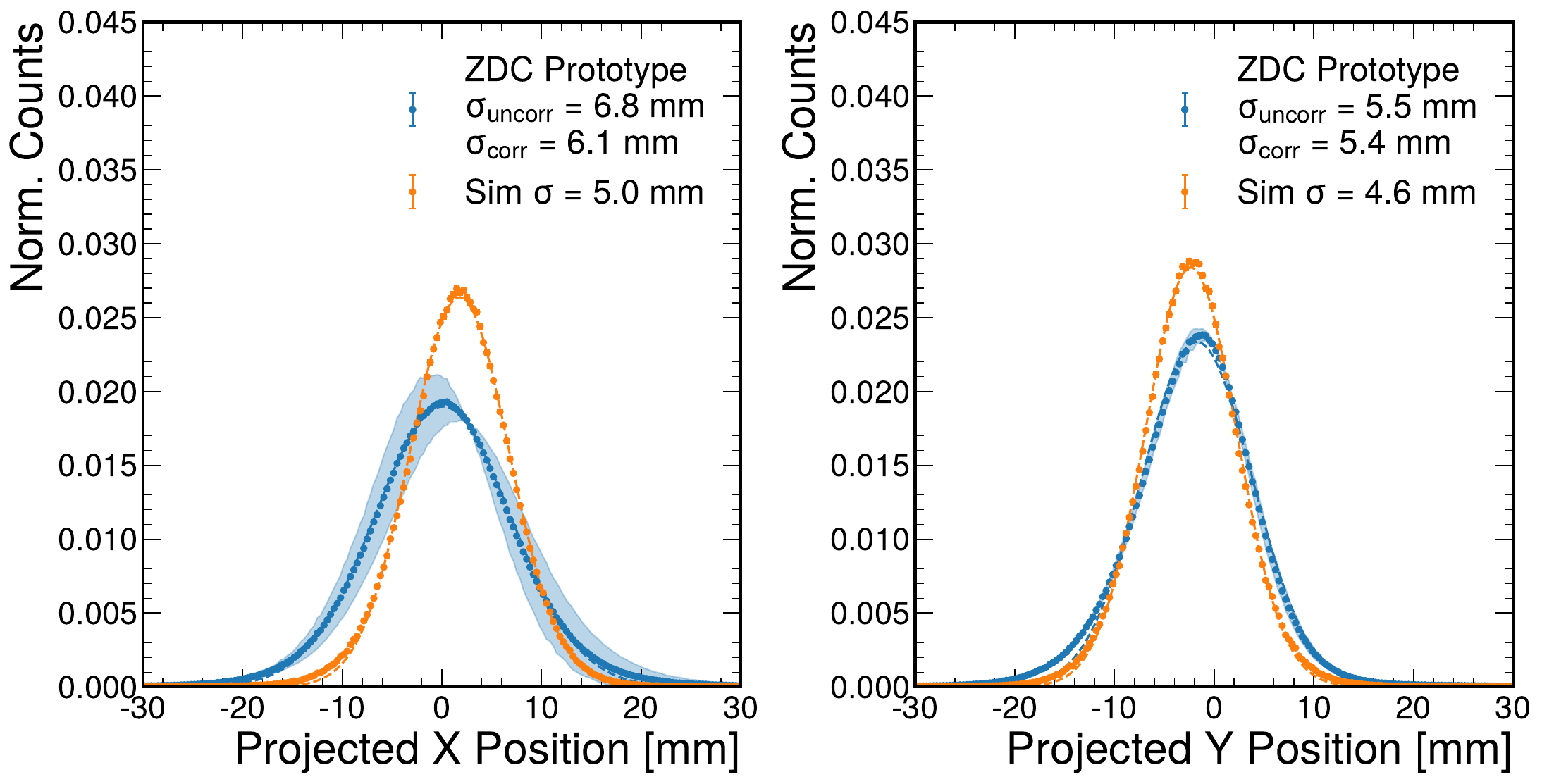}
        \caption{Position distributions of the principle axes of particle showers projected to the front surface of the detector, compared to a simulation of 5.3 GeV positrons. The combined systematic uncertainty is represented as an error band, and statistical uncertainty is represented with error bars.}
        \label{fig:position_res}
\end{figure}

After these corrections, the position resolution in $x$ is found to be $\sigma = 6.1$~mm in data and $\sigma = 5.0$~mm in simulation, while in $y$ it is $\sigma = 5.4$~mm in data and $\sigma = 4.6$~mm in simulation, shown in Fig. \ref{fig:position_res}. 

\begin{figure}[h!]
    \centering
        \includegraphics[width=0.99\linewidth]{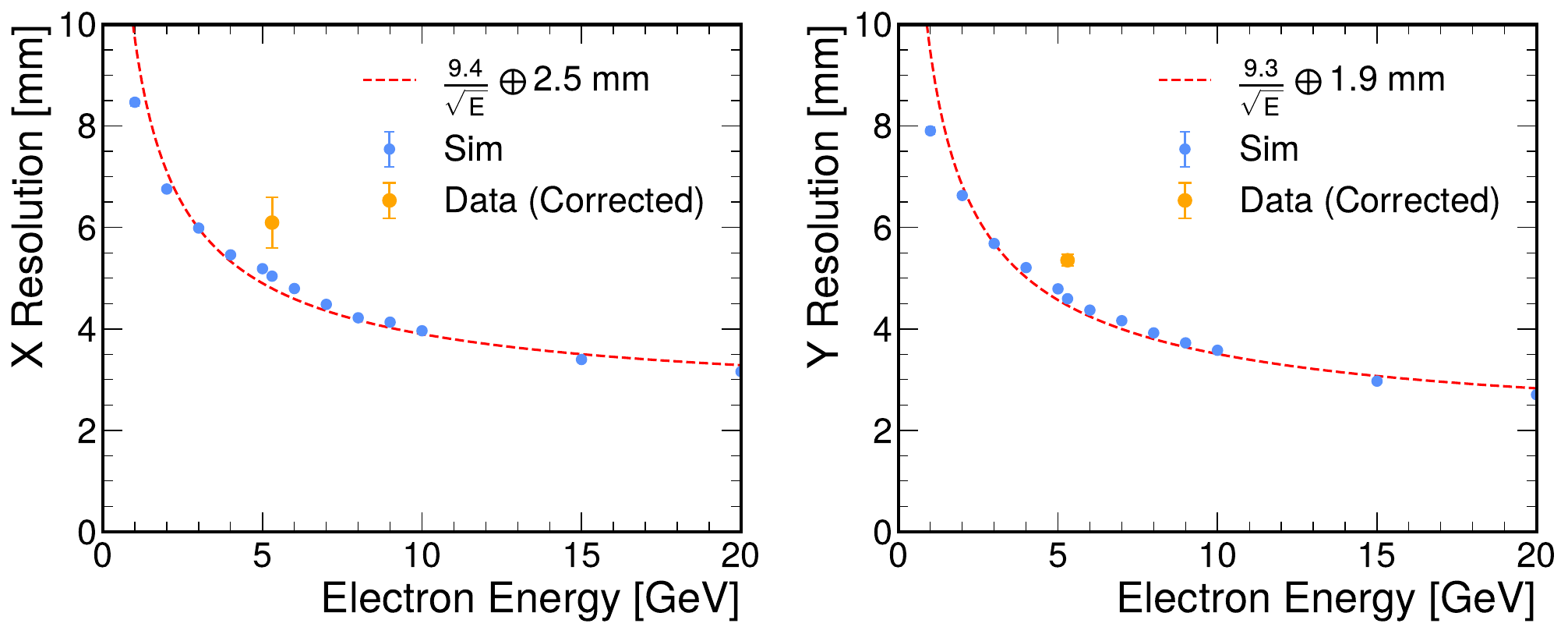}
        \caption{Simulated position resolutions in $x$ and $y$ as a function of particle energy from 1–20~GeV. The measured and corrected position resolution at 5.3~GeV is shown as a single data point. The error bars on this point represent the combined statistical and systematic uncertainties, added in quadrature and symmetrized.}
        \label{fig:position_res_e}
\end{figure}

Figure~\ref{fig:position_res_e} shows the simulated position resolutions in $x$ and $y$ across electron energies from 1–20~GeV. In generating these points, the 1~GeV event energy threshold was not applied in the simulation. This indicates the expected energy dependence of the position resolution for the ZDC prototype. The measured and corrected resolution at 5.3~GeV is included as a single reference point.

\subsubsection{Energy Distributions}
\begin{figure}[h!]
    \centering
        \includegraphics[width=0.7\linewidth]{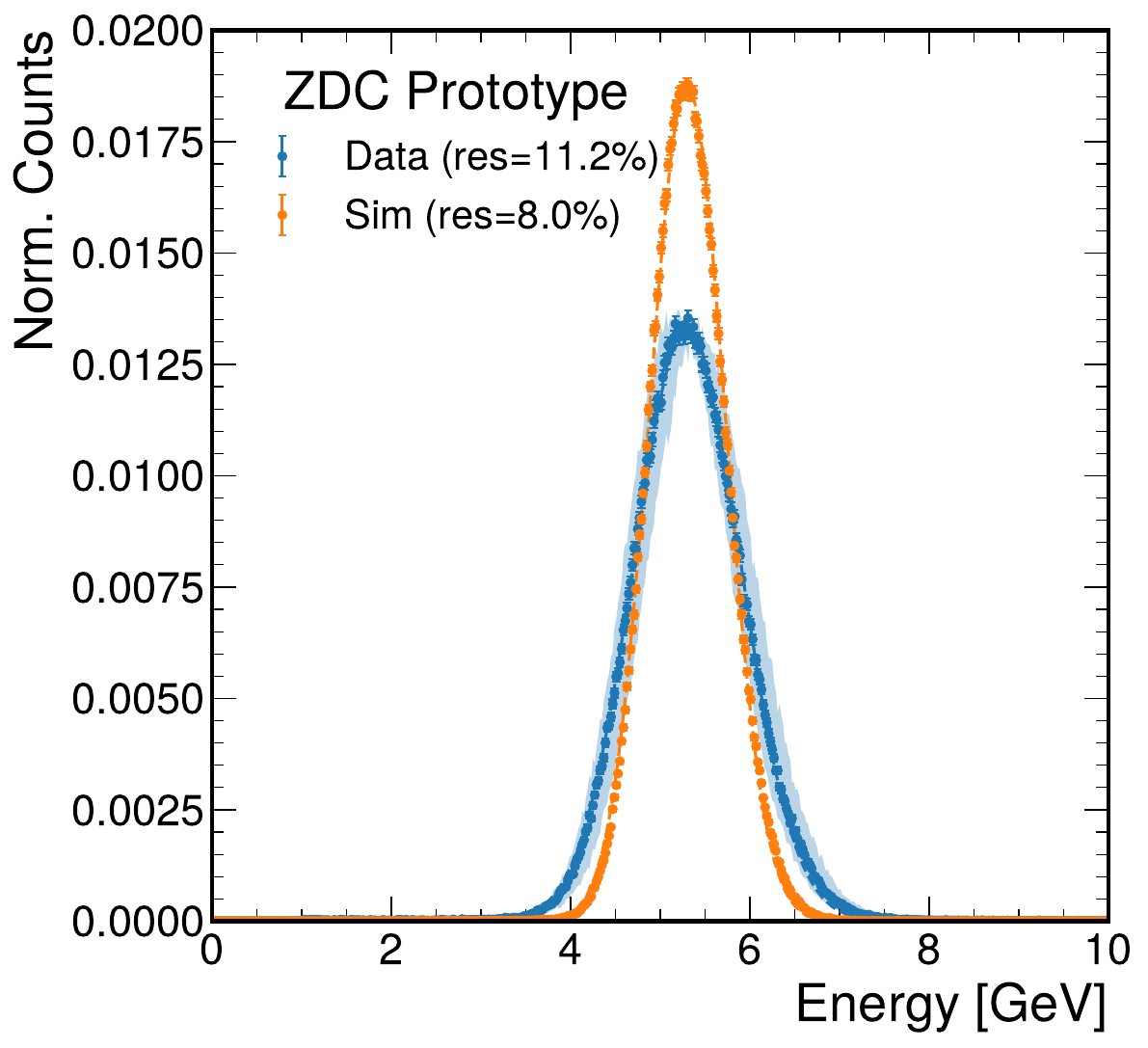}
        \caption{Event energy compared to a simulation of 5.3 GeV positrons. The combined systematic uncertainty is represented as an error band, and statistical uncertainty is represented with error bars.}
        \label{fig:event_energy}
\end{figure}

The event energy spectra are shown in Fig.~\ref{fig:event_energy}. In data, the measured resolution includes contributions from the positron energy spread of the Hall~D PS~\cite{somov2025private} and from the trigger tile width, each contributing approximately 25~MeV. Combined in quadrature, these effects produce a contribution that was subtracted from the measured resolution to obtain the intrinsic detector energy resolution. This correction has a negligible effect, leaving the energy resolution at 11.1\%. The simulation using 5.3~GeV positrons yields an energy resolution of 8.1\%, with no such beam energy spread included.

\begin{figure}[h!]
    \centering
        \includegraphics[width=0.7\linewidth]{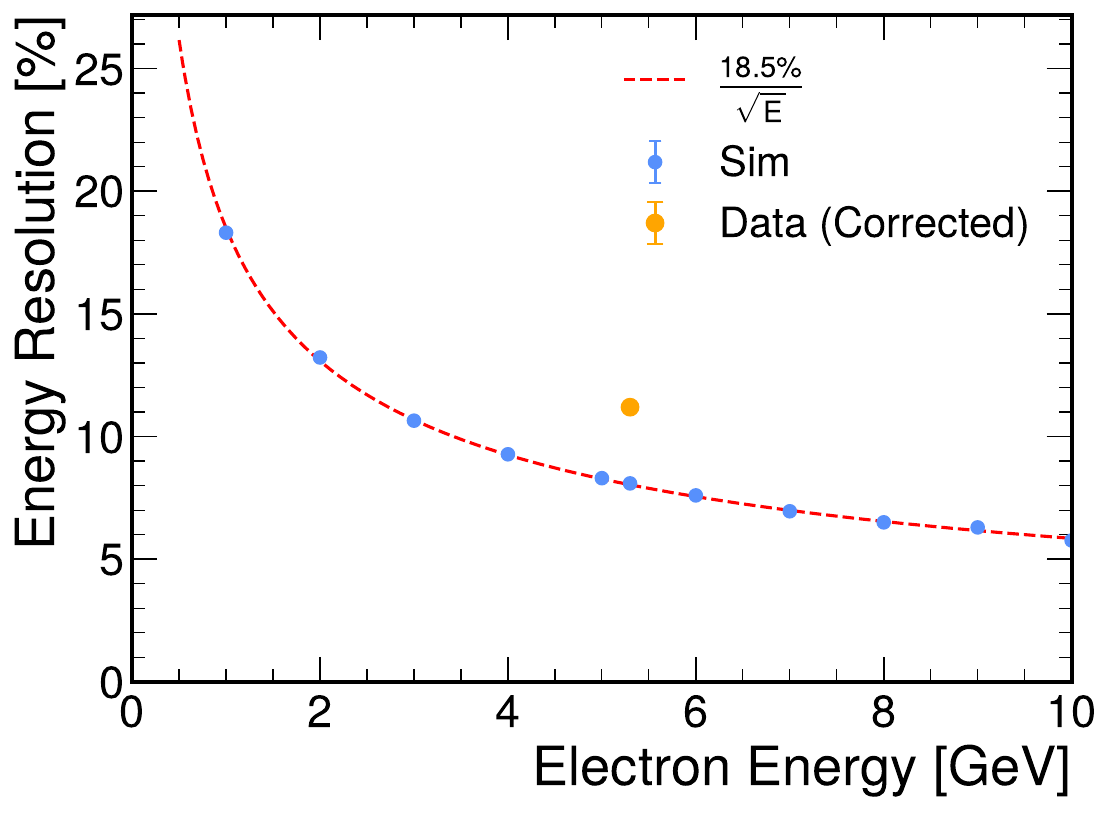}
        \caption{Simulated energy resolutions from 1–10~GeV. The measured energy resolution at 5.3~GeV is shown as a single data point with combined statistical and systematic uncertainties, which are smaller than the marker size.}
        \label{fig:energy_res}
\end{figure}

Figure~\ref{fig:energy_res} presents the simulated energy resolutions as a function of electron energy from 1–10~GeV. For this analysis, the 1~GeV event energy cut was omitted in the simulation. The simulation points show how the energy resolution is expected to vary with particle energy for the ZDC prototype. The measured energy resolution at 5.3~GeV is shown as a single data point for comparison.

\begin{figure}[h!]
    \centering
        \includegraphics[width=0.99\linewidth]{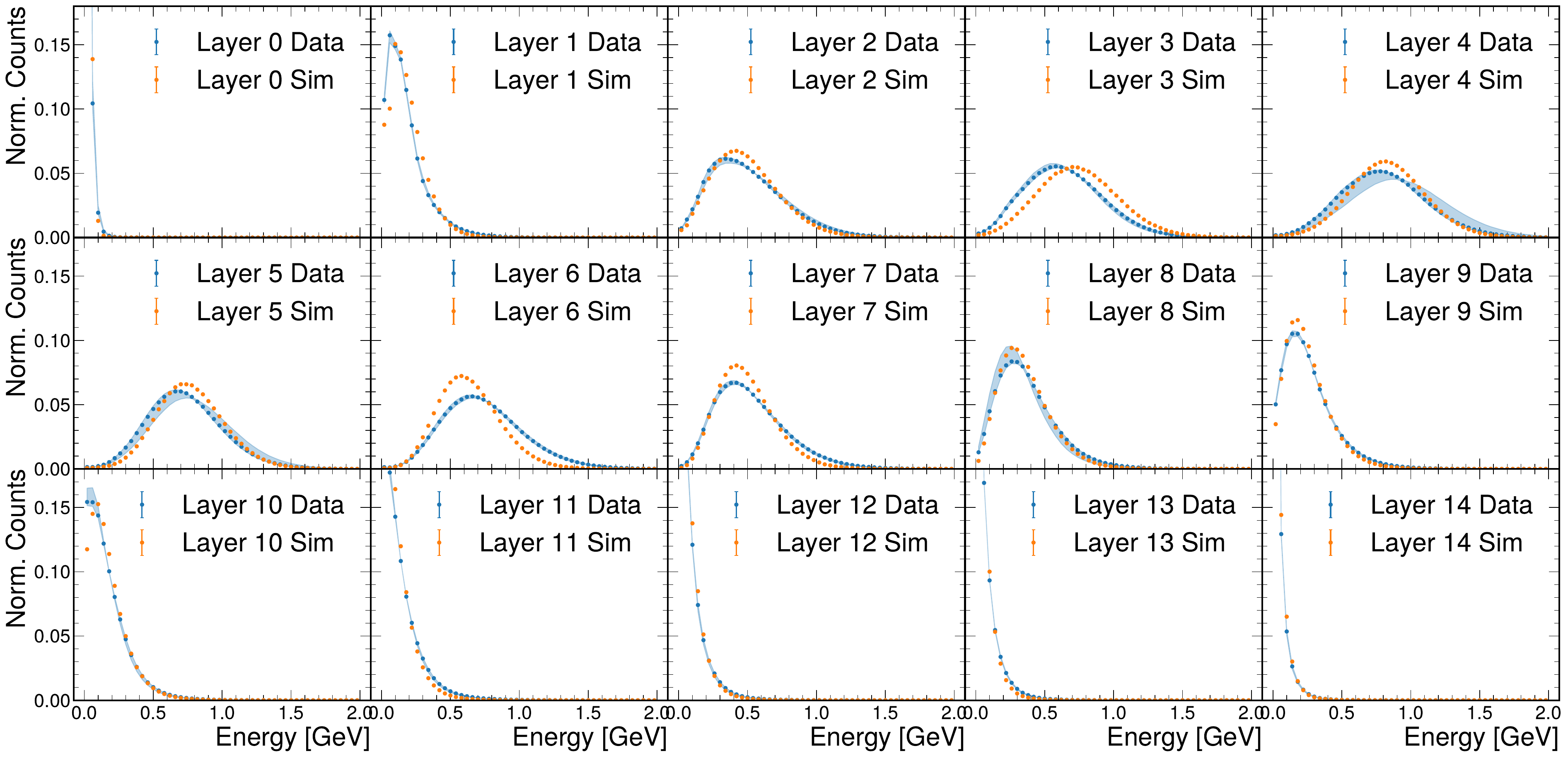}
        \caption{Event layer energy distributions compared to a simulation of 5.3 GeV positrons. The combined systematic uncertainty is represented as an error band, and statistical uncertainty is represented with error bars.} 
        \label{fig:layer_energy}
\end{figure}

\begin{figure}[h!]
    \centering
        \includegraphics[width=0.8\linewidth]{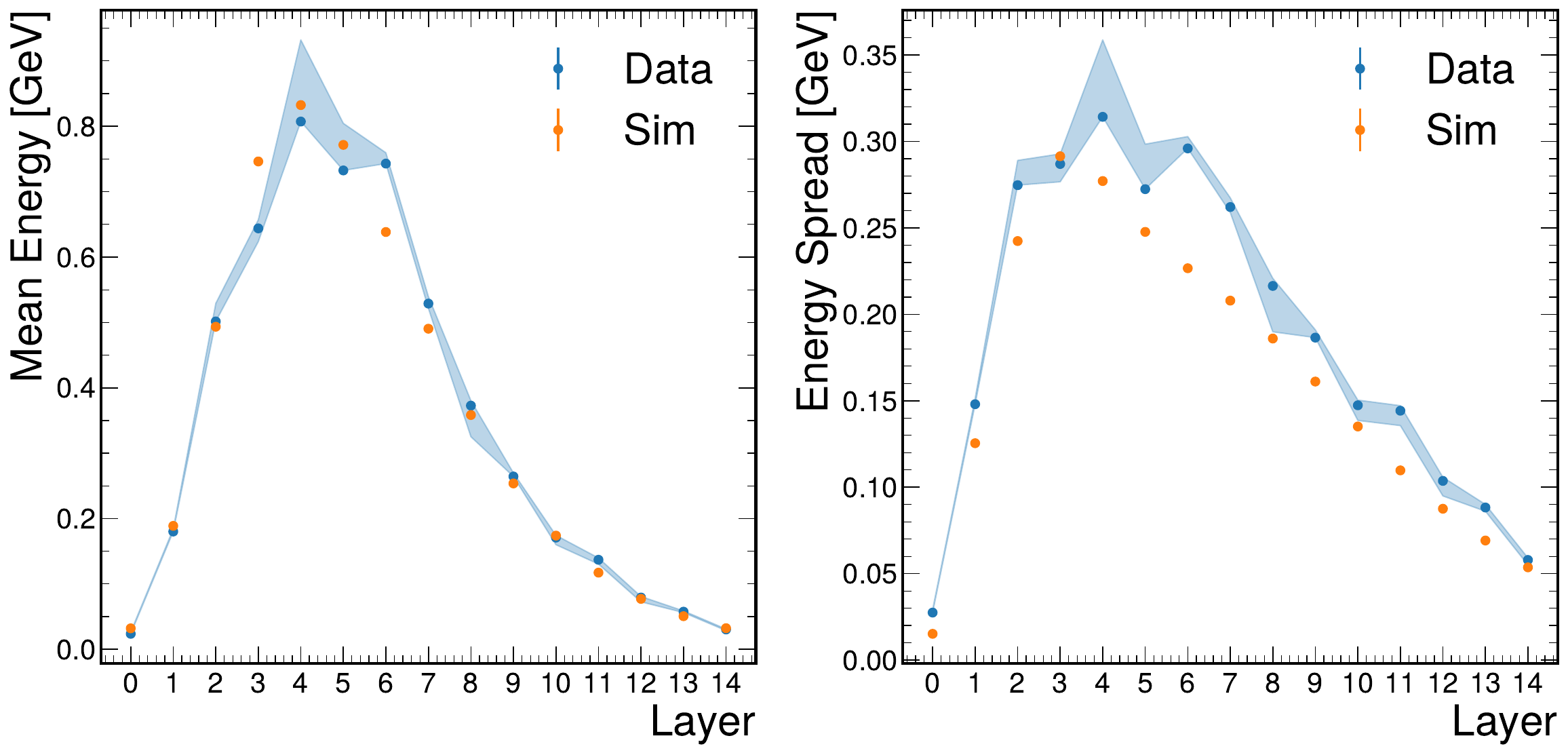}
        \caption{Average layer event energy per layer (left), and the standard deviation of layer event energy per layer (right), both compared to a simulation of 5.3 GeV positrons. The combined systematic uncertainty is represented as an error band, and statistical uncertainty is represented with error bars.}
        \label{fig:layer_energy_summary}
\end{figure}

The summed event energy distributions for each layer, shown in Fig.~\ref{fig:layer_energy} and summarized in Fig.~\ref{fig:layer_energy_summary}, provide the detector response as a function of depth and characterize the longitudinal shower shape.

\FloatBarrier
\section{Summary and Conclusions}
\label{sec:conclusions}

We have presented the first beam test of a SiPM-on-tile Zero Degree Calorimeter (ZDC) prototype, conducted at Jefferson Laboratory using 5.3 GeV positrons. The detector comprises 370 readout channels and implements the staggered scintillator geometry planned for the final ePIC ZDC design, corresponding to $\mathcal{O}(10\%)$ of the full ZDC channel count and representing a significant advancement over previous prototypes.

Key performance characteristics of the ZDC prototype were evaluated and compared with GEANT4 simulations. Measurements of shower shape and spatial reconstruction indicate that the overall detector response and main features observed in the data are reasonably reproduced by the simulation. Differences in the energy response are attributed to detector effects not fully captured in the model, such as tile non-uniformity and other instrumental variations.

These results establish a strong experimental foundation for the ePIC ZDC and provide input for the optimization of assembly methods, calibration procedures, and detector layout. The successful operation and performance of a SiPM-on-tile sampling calorimeter at this scale demonstrate the feasibility of bridging the gap between earlier small-scale tests and the full detector implementation.

\appendix

\section*{Code and Data Availability}
The dataset used in this study is publicly available on Zenodo~\cite{Preins_2026_ZDC_Prototype_JLab_Test_Data}. 
The simulation and analysis scripts are available on GitHub~\cite{SeanBP_2026_ZDC_JLab_Analysis}.

\subsection*{Contributions}
Sean Preins was the primary data analyst and contributed to construction, bench-top quality assurance tests, installation, operation, and editing of the paper. Weibin Zhang developed the event building script used in the analysis. Ryan Tsiao, Mia Macias, and Brice Saunders contributed to the construction and installation of the test module. Love Preet assisted in the installation of the test module. Miguel Arratia conceived, supervised, and edited this work.

\acknowledgments
We'd like to thank Andrew Lumanong for helping our group coordinate installation of the detector in Hall-D and assist with its alignment. We would also like to thank Sasha Somov for valuable insight into the beam conditions of the Hall-D Pair Spectrometer. We thank members of the California EIC consortium, the ePIC collaboration, and in particular Yulia Furletova for supporting this project.

\bibliographystyle{utphys} 
\bibliography{biblio.bib}

@Article{Accardi2016,
author={Accardi, A. and others},
title={Electron-Ion Collider: The next QCD frontier},
journal={The European Physical Journal A},
year={2016},
month={Sep},
day={08},
volume={52},
number={9},
pages={268},
issn={1434-601X},
doi={10.1140/epja/i2016-16268-9},
url={https://doi.org/10.1140/epja/i2016-16268-9}
}

@article{ABDULKHALEK2022122447,
title = {Science Requirements and Detector Concepts for the Electron-Ion Collider: EIC Yellow Report},
journal = {Nuclear Physics A},
volume = {1026},
pages = {122447},
year = {2022},
issn = {0375-9474},
doi = {https://doi.org/10.1016/j.nuclphysa.2022.122447},
url = {https://www.sciencedirect.com/science/article/pii/S0375947422000677},
author = {R. {Abdul Khalek} and others}
}

@misc{ePIC,
  author       = {{ePIC Collaboration}},
  title        = {The ePIC Detector},
  howpublished = {\url{https://www.bnl.gov/eic/epic.php}},
  note         = {Accessed: 2025-11-01},
  year         = {2025}
}

@article{MILTON2025170613,
title = {Design and simulation of a SiPM-on-tile ZDC for the future EIC, and its performance with graph neural networks},
journal = {Nuclear Instruments and Methods in Physics Research Section A: Accelerators, Spectrometers, Detectors and Associated Equipment},
volume = {1079},
pages = {170613},
year = {2025},
issn = {0168-9002},
doi = {https://doi.org/10.1016/j.nima.2025.170613},
url = {https://www.sciencedirect.com/science/article/pii/S0168900225004140},
author = {Ryan Milton and Sebouh J. Paul and Barak Schmookler and Miguel Arratia and Piyush Karande and Aaron Angerami and Fernando {Torales Acosta} and Benjamin Nachman}
}

@article{Arratia_2023,
doi = {10.1088/1748-0221/18/05/P05045},
url = {https://dx.doi.org/10.1088/1748-0221/18/05/P05045},
year = {2023},
month = {may},
publisher = {IOP Publishing},
volume = {18},
number = {05},
pages = {P05045},
author = {Arratia, Miguel and Garabito Ruiz, Luis and Huang, Jiajun and Paul, Sebouh J. and Preins, Sean and Rodriguez, Miguel},
title = {Studies of time resolution, light yield, and crosstalk using SiPM-on-tile calorimetry for the future Electron-Ion Collider},
journal = {Journal of Instrumentation}
}

@article{ARRATIA2023167866,
title = {A high-granularity calorimeter insert based on SiPM-on-tile technology at the future Electron-Ion Collider},
journal = {Nuclear Instruments and Methods in Physics Research Section A: Accelerators, Spectrometers, Detectors and Associated Equipment},
volume = {1047},
pages = {167866},
year = {2023},
issn = {0168-9002},
doi = {https://doi.org/10.1016/j.nima.2022.167866},
url = {https://www.sciencedirect.com/science/article/pii/S0168900222011585},
author = {Miguel Arratia and Kenneth Barish and Liam Blanchard and Huan Z. Huang and Zhongling Ji and Bishnu Karki and Owen Long and Ryan Milton and Ananya Paul and Sebouh J. Paul and Sean Preins and Barak Schmookler and Oleg Tsai and Zhiwan Xu}
}

@article{Paul_2024,
title={Leveraging staggered tessellation for enhanced spatial resolution in high-granularity calorimeters},
volume={1060},
ISSN={0168-9002},
url={http://dx.doi.org/10.1016/j.nima.2023.169044},
DOI={10.1016/j.nima.2023.169044},
journal={Nuclear Instruments and Methods in Physics Research Section A: Accelerators, Spectrometers, Detectors and Associated Equipment},
publisher={Elsevier BV},
author={Paul, Sebouh J. and Arratia, Miguel},
year={2024},
month=mar, pages={169044} 
}

@article{instruments7040043,
AUTHOR = {Arratia, Miguel and Bagby, Bruce and Carney, Peter and Huang, Jiajun and Milton, Ryan and Paul, Sebouh J. and Preins, Sean and Rodriguez, Miguel and Zhang, Weibin},
TITLE = {Beam Test of the First Prototype of SiPM-on-Tile Calorimeter Insert for the EIC Using 4 GeV Positrons at Jefferson Laboratory},
JOURNAL = {Instruments},
VOLUME = {7},
YEAR = {2023},
NUMBER = {4},
ARTICLE-NUMBER = {43},
URL = {https://www.mdpi.com/2410-390X/7/4/43},
ISSN = {2410-390X},
DOI = {10.3390/instruments7040043}
}

@article{Zhang_2025,
doi = {10.1088/1748-0221/20/06/P06029},
url = {https://dx.doi.org/10.1088/1748-0221/20/06/P06029},
year = {2025},
month = {jun},
publisher = {IOP Publishing},
volume = {20},
number = {06},
pages = {P06029},
author = {Zhang, Weibin and Preins, Sean and Huang, Jiajun and Paul, Sebouh J. and Milton, Ryan and Rodriguez, Miguel and Carney, Peter and Tsiao, Ryan and Abdelkadous, Yousef and Arratia, Miguel},
title = {First-ever deployment of a SiPM-on-tile calorimeter in a collider: a parasitic test with 200 GeV p p collisions at RHIC.},
journal = {Journal of Instrumentation}
}

@article{Zhang:2025cmc,
    author = "Zhang, Weibin and Liang, Xilin and Preins, Sean and Arratia, Miguel",
    title = "{Calibration of an Irradiated Prototype for the EIC Zero-Degree Calorimeter}",
    eprint = "2512.20852",
    archivePrefix = "arXiv",
    primaryClass = "physics.ins-det",
    month = "12",
    year = "2025"
}

@misc{UCDavisTest,
      title={Measurement of SiPM Dark Currents and Annealing Recovery for Fluences Expected in ePIC Calorimeters at the Electron-Ion Collider}, 
      author={Jiajun Huang and Sean Preins and Ryan Tsiao and Miguel Rodriguez and Barak Schmookler and Miguel Arratia},
      year={2025},
      eprint={2503.14622},
      archivePrefix={arXiv},
      primaryClass={physics.ins-det},
      url={https://arxiv.org/abs/2503.14622}, 
}

@article{BARBOSA2015376,
title = {Pair spectrometer hodoscope for Hall D at Jefferson Lab},
journal = {Nuclear Instruments and Methods in Physics Research Section A: Accelerators, Spectrometers, Detectors and Associated Equipment},
volume = {795},
pages = {376-380},
year = {2015},
issn = {0168-9002},
doi = {https://doi.org/10.1016/j.nima.2015.06.012},
url = {https://www.sciencedirect.com/science/article/pii/S0168900215007573},
author = {F. Barbosa and C. Hutton and A. Sitnikov and A. Somov and S. Somov and I. Tolstukhin}
}

@misc{somov2025private,
  author       = {S. Somov},
  title        = {Private communication on the beam conditions of the Hall D Pair Spectrometer},
  howpublished = {Private communication},
  note         = {Accessed September 9, 2025}
}

@dataset{Preins_2026_ZDC_Prototype_JLab_Test_Data,
  author       = {Preins, Sean and Arratia, Miguel},
  title        = {ZDC Prototype JLab Test Data},
  year         = {2026},
  month        = feb,
  publisher    = {Zenodo},
  doi          = {10.5281/zenodo.18726771},
  url          = {https://doi.org/10.5281/zenodo.18726771},
  note         = {Data collected at the Thomas Jefferson National Laboratory Hall D Pair Spectrometer for an EIC ZDC prototype detector}
}

@software{SeanBP_2026_ZDC_JLab_Analysis,
  author       = {Preins, Sean},
  title        = {ZDC-JLab-Analysis},
  year         = {2026},
  publisher    = {GitHub},
  url          = {https://github.com/SeanBP/ZDC-JLab-Analysis},
  note         = {Simulation and analysis scripts for the ZDC prototype JLab test}
}

@article{BOCK2023168464,
title = {Design and simulated performance of calorimetry systems for the ECCE detector at the electron ion collider},
journal = {Nuclear Instruments and Methods in Physics Research Section A: Accelerators, Spectrometers, Detectors and Associated Equipment},
volume = {1055},
pages = {168464},
year = {2023},
issn = {0168-9002},
doi = {https://doi.org/10.1016/j.nima.2023.168464},
url = {https://www.sciencedirect.com/science/article/pii/S0168900223004540},
author = {F. Bock and others}
}

@article{RevModPhys.88.015003,
  title = {Experimental tests of particle flow calorimetry},
  author = {Sefkow, Felix and White, Andy and Kawagoe, Kiyotomo and P\"oschl, Roman and Repond, Jos\'e},
  journal = {Rev. Mod. Phys.},
  volume = {88},
  issue = {1},
  pages = {015003},
  numpages = {53},
  year = {2016},
  month = {Feb},
  publisher = {American Physical Society},
  doi = {10.1103/RevModPhys.88.015003},
  url = {https://link.aps.org/doi/10.1103/RevModPhys.88.015003}
}

@misc{CERN:HL-LHC,
  title        = {The High-Luminosity LHC},
  howpublished = {\url{https://cds.cern.ch/record/2114693}}
}

@techreport{Contardo:2015_phase2,
  author       = {Contardo, D. and others},
  title        = {Technical Proposal for the Phase-II Upgrade of the CMS Detector},
  reportNumber = {CERN-LHCC-2015-010},
  doi          = {10.17181/CERN.VU8I.D59J}
}

@techreport{CMS:2017_HGCAL_TDR,
  author       = {{CMS Collaboration}},
  title        = {The Phase-2 Upgrade of the CMS Endcap Calorimeter},
  reportNumber = {CERN-LHCC-2017-023},
  doi          = {10.17181/CERN.IV8M.1JY2}
}

@article{THOMSON200925,
title = {Particle flow calorimetry and the PandoraPFA algorithm},
journal = {Nuclear Instruments and Methods in Physics Research Section A: Accelerators, Spectrometers, Detectors and Associated Equipment},
volume = {611},
number = {1},
pages = {25-40},
year = {2009},
issn = {0168-9002},
doi = {https://doi.org/10.1016/j.nima.2009.09.009},
url = {https://www.sciencedirect.com/science/article/pii/S0168900209017264},
author = {M.A. Thomson},
}

@misc{theildconceptgroup2010internationallargedetectorletter,
      title={The International Large Detector: Letter of Intent}, 
      author={The ILD Concept Group},
      year={2010},
      eprint={1006.3396},
      archivePrefix={arXiv},
      primaryClass={hep-ex},
      url={https://arxiv.org/abs/1006.3396}, 
}

@misc{theildcollaboration2020internationallargedetectorinterim,
      title={International Large Detector: Interim Design Report}, 
      author={The ILD Collaboration},
      year={2020},
      eprint={2003.01116},
      archivePrefix={arXiv},
      primaryClass={physics.ins-det},
      url={https://arxiv.org/abs/2003.01116}, 
}

@misc{linssen2012physicsdetectorsclicclic,
      title={Physics and Detectors at CLIC: CLIC Conceptual Design Report}, 
      author={Lucie Linssen and Akiya Miyamoto and Marcel Stanitzki and Harry Weerts},
      year={2012},
      eprint={1202.5940},
      archivePrefix={arXiv},
      primaryClass={physics.ins-det},
      url={https://arxiv.org/abs/1202.5940}, 
}

@article{FCC_ee_CDR_Vol2,
  author = {{FCC collaboration}},
  title = {{FCC-ee: The Lepton Collider: Future Circular Collider Conceptual Design Report Volume 2}},
  journal = {Eur. Phys. J. ST},
  volume = {228},
  year = {2019},
  pages = {261},
  doi = {10.1140/epjst/e2019-900045-4}
}

@misc{thecepcstudygroup2018cepcconceptualdesignreport,
      title={CEPC Conceptual Design Report: Volume 2 - Physics \& Detector}, 
      author={The CEPC Study Group},
      year={2018},
      eprint={1811.10545},
      archivePrefix={arXiv},
      primaryClass={hep-ex},
      url={https://arxiv.org/abs/1811.10545}, 
}

\end{document}